\documentclass{article}

\usepackage{microtype}
\usepackage{graphicx}
\usepackage{subcaption}  
\usepackage{textcomp} 
\usepackage{booktabs} 
\usepackage{amsthm} 
\usepackage{amsmath}
\usepackage{svg}
\newtheorem{hypothesis}{Hypothesis}
\usepackage{hyperref}

\usepackage[accepted]{icml2021}

\icmltitlerunning{Alpha Mining and Enhancing via Warm Start Genetic Programming for Quantitative Investment}

\begin{document}

\twocolumn[
\icmltitle{Alpha Mining and Enhancing via Warm Start Genetic Programming for Quantitative Investment}



\icmlsetsymbol{equal}{*}

\begin{icmlauthorlist}
\icmlauthor{Weizhe Ren}{to1}
\icmlauthor{Yichen Qin}{to2}
\icmlauthor{Yang Li}{to1}
\end{icmlauthorlist}

\icmlaffiliation{to1}{Center for Applied Statistics and School of Statistics, Renmin University of China, Beijing, China}
\icmlaffiliation{to2}{Department of Operations, Business Analytics, and Information Systems, University of Cincinnati, Cincinnati, OH}

\icmlcorrespondingauthor{Yang Li}{yang.li@ruc.edu.cn}

\icmlkeywords{Stock selection Factors; Genetic Programming; Portfolio; Stock Prediction}

\vskip 0.3in
]

\printAffiliationsAndNotice{}

\begin{abstract}
Traditional genetic programming (GP) often struggles in stock alpha factor discovery due to its vast search space, overwhelming computational burden, and sporadic effective alphas. 
We find that GP performs better when focusing on promising regions rather than random searching. 
This paper proposes a new GP framework with carefully chosen initialization and structural constraints to enhance search performance and improve the interpretability of the alpha factors.
This approach is motivated by  and mimics the alpha searching practice and aims to boost the efficiency of such a process.
Analysis of 2020-2024 Chinese stock market data shows that our method yields superior out-of-sample prediction results and higher portfolio returns than the benchmark.

\end{abstract}

\section{Introduction}
Predicting future stock returns is one of the most challenging tasks in quantitative trading. 
Stock prices are influenced by numerous factors, such as company performance, investor sentiment, government policies, and other relevant variables. 
To explain stock market fluctuations, economists have developed a variety of theoretical models, including the Capital Asset Pricing Model (CAPM) \cite{Sharpe64}, the Fama-French three-factor model \cite{FamaFrench93}, the Fama-French five-factor model \cite{FamaFrench15}, and others. 
In quantitative trading practice, designing new factors that can explain and predict asset returns is crucial to the profitability of a strategy. 
Such factors are typically referred to as alphas, alpha factors or stock selection factors. 
\citet{Narang13} defines stock selection as "the process of using data-driven methods to build models, select stocks that are expected to perform well in the future, and generate returns." 
Based on this understanding, the value of an alpha factor of a given stock on a specific trading day can be seen as a prediction of its future returns: the higher the value, the more likely the stock is to achieve relatively higher returns in the future.

Traditionally, new alphas have typically been generated through theoretical hypotheses proposed by economists or financial engineers, which are then converted into mathematical formulas and subsequently validated using historical data. 
These manually constructed alphas generally exhibit strong interpretability and stability.
However, they are often unable to capture complex nonlinear relationships within the data.
Moreover, the manual construction process is not only time-consuming but also requires extensive investment experience \cite{Feng19}, making it difficult to generalize to the broader investing community \cite{Becker09}.

With the advent of the big data era and advancements in computational power, both academia and the financial industry have gradually shifted towards data-driven, automated alpha discovery methods, particularly those based on trading data, such as trading volume, transaction prices, turnover rate, etc, to extract more complex market information. 
Such data is often more real-time and can reflect the dynamic characteristics of stock price movements, from which more accurate and forward-looking alphas can often be uncovered \cite{Cui21}.

Among the automated methods for alpha discovery, genetic programming-based method is one of the most popular approaches. 
Genetic programming (GP) can yield formulaic factors based on simple mathematical structures, which are more interpretable, easier to understand and apply, and can be modified promptly. 
Therefore, how to use GP to derive stock selection factors with good performance has become a key focus for many economists and financial researchers.

\begin{figure}[ht]
\vskip 0.2in
\begin{center}
\centerline{\includegraphics[width=0.6\columnwidth]{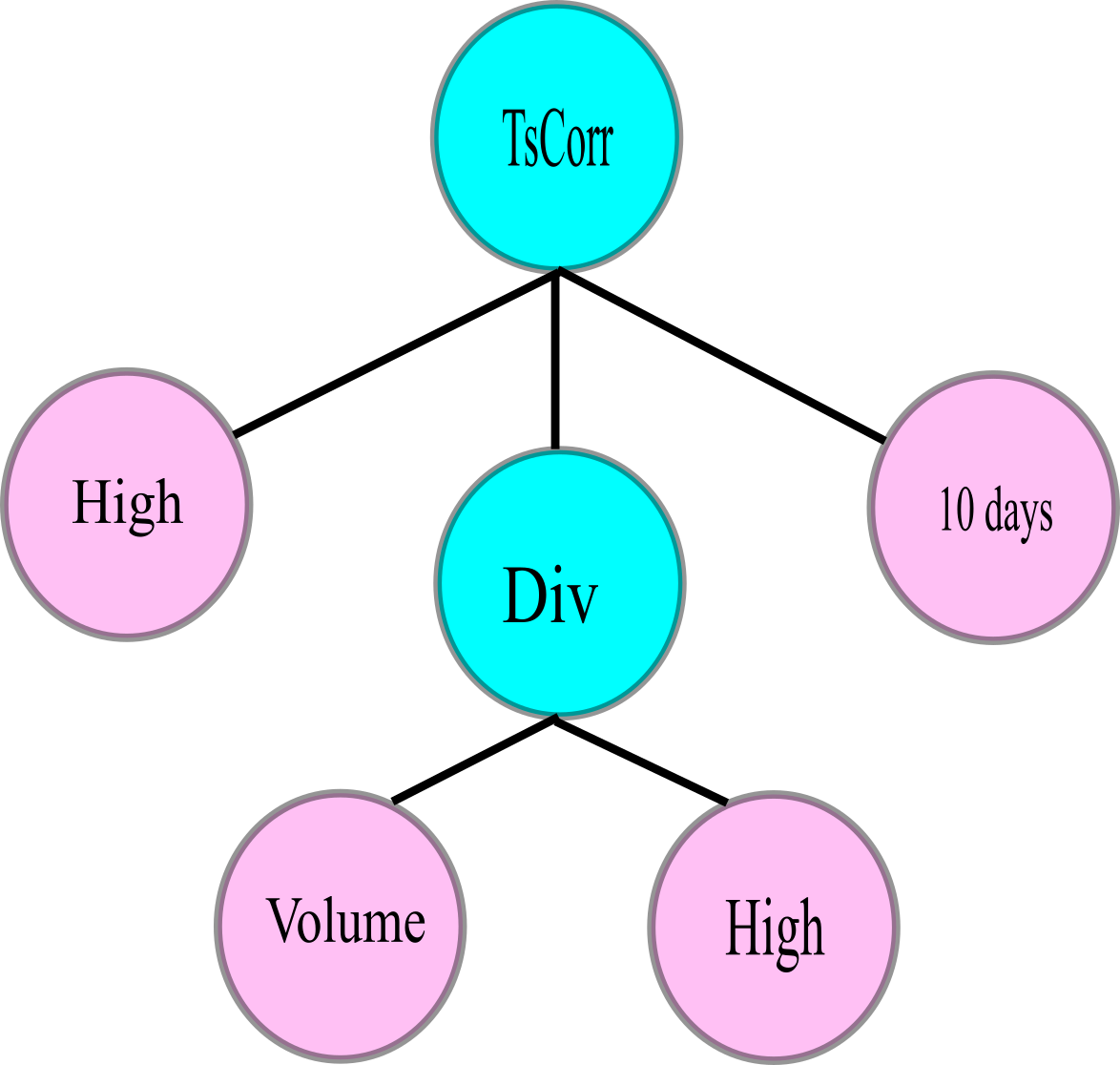}}
\caption{An example of a GP tree. The green parts are root nodes and the pink parts are leaf nodes.}
\label{fig:GP tree}
\end{center}
\vskip -0.2in
\end{figure}

Genetic Programming is a branch of Genetic Algorithms (GA). 
In GP, solutions are represented in a tree-like structure, where the root node represents operators and the leaf nodes represent input data or variables, as shown in the Figure \ref{fig:GP tree}. 
This structure enables GP to generate complex programs or formulas \cite{Koza92}. 
Like GA, GP optimizes the target solutions using the concept of biological evolution. 
In the context of alpha discovery, traditional GP begins by randomly generating an initial population of factors. 
Starting from this initial population, GP uses specific rules to select individuals. 
The selected individuals undergo crossover or mutation to produce offspring (similar to chromosome crossover and mutation in genetic evolution), resulting in a new population with new genetic traits. 
The process continues by updating the individuals' fitness values to identify the best solution \cite{Xia06}. 

In the field of mining stock selection factors, the most commonly used fitness measure is the Information Coefficient (IC), which can reflect the stock selection ability of a single alpha. 
The calculation formula for IC is as follows, where the IC of an alpha is obtained by averaging the daily cross-sectional IC values over a period of time, and the cross-sectional IC represents the correlation between the alpha values and future returns on a given day.
\footnote{We assume all IC metrics are $\geq 0$, as negative IC values can be flipped by multiplying the alpha by -1.}
\[
IC = \frac{1}{T} \sum_{t=1}^{T} \text{PearsonCorr}(\textbf{a}_t, \textbf{r}_{t})
\]
where $\textbf{a}_t$ is the value vector of the alpha at date $t$, $\textbf{r}_{t}$ is the value vector of the forward returns at date $t$.
In this paper, we use the five-day VWAP
\footnote{VWAP is short for the Volume Weighted Average Price.}
cumulative return as our future return, where $\textbf{r}_t =\frac{\text{VWAP}_{t+6}}{\text{VWAP}_{t+1}} - 1$.

As early as 1995, \citet{Allen99} introduced GP into the field of stock investment. 
In 2016, the quantitative investment management firm WorldQuant released a report titled "WorldQuant: Formulaic 101 Alphas," in which they disclosed 101 formulaic alphas derived using GP \cite{Kakushadze16}. 
In 2017, Guotai Junan Securities used GP to construct 191 short-term stock selection factors and applied them to build a multi-factor stock selection strategy \cite{GuoTai2017}. 
To this day, a large amount of quantitative trading still relies on GP to mine formulaic alphas for predicting stock trends \cite{ShenWan2016, HuaTai2019_1, HuaTai2019_2, GuoTai2023, HuaTai2024}.

\section{Problem Statement}
As research and applications increasingly focus on using GP to mine alphas for stock selection, several challenges associated with this method have emerged. 
\citet{ONeill2010} raised some major open issues and development directions that need to be addressed in general  GP. 
\citet{Brabazon2020} discussed the significant potential of GP in the financial domain but highlighted that its applications are predominantly experimental implementations of computer science on financial data, with limited exploration in financial literature. 
They further analyzed issues related to GP’s scalability, benchmarking, and data snooping, offering explanations and future prospects. 
\citet{Kim2008} proposed a constrained tree structure to reduce the risk of overfitting in GP. 
\citet{Long2022} identified two advanced methods that could improve the application of GP in alpha mining: multi-objective GP to simulate trade-offs in real-world trading and tree structure constraints to mitigate overfitting risks. 
\citet{Gupta2012} highlighted the premature convergence problem in genetic algorithms, where a single effective gene dominates the population, reducing diversity and trapping the population in suboptimal states. 
Similarly, \citet{Zhang2020} identified three challenges in using GP for alpha mining: rapidly identifying promising search spaces, guiding searches away from already explored areas, and preventing premature convergence.

\begin{figure}[ht]
\vskip 0.2in
\begin{center}
\centerline{\includegraphics[width=\columnwidth]{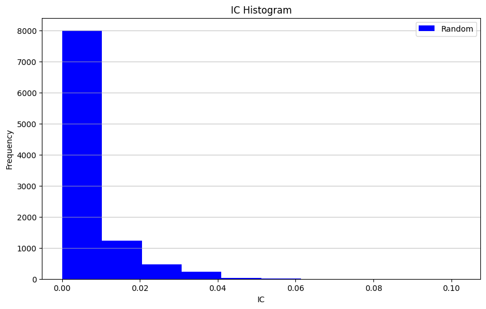}}
\caption{The results indicate that effective alphas are sparse within the search space of traditional GP.}
\label{fig:Random Factor1}
\end{center}
\vskip -0.2in
\end{figure}

In our implementation of traditional GP for alpha mining, the most significant challenges are the infinite search space and the sparsity of effective alphas. 
The infinite search space arises from the unrestricted growth in the depth and length of tree-structured alphas. 
On the other hand, even manually designing effective stock selection factors is a non-trivial task, indicating that effective alphas are naturally sparse within such an enormous search space. 
To better understand this sparsity, we design an experiment to validate it. 
Using traditional GP, we randomly generate 10,000 alphas on a little test data set and find that fewer than $3\%$ are effective (with an $IC > 0.03$ \footnote{In this experiment, we assume an alpha with an IC greater than 0.03 is considered effective.}), while the vast majority of alphas had IC values close to 0, as shown in the Figure \ref{fig:Random Factor1}. 
This result more intuitively illustrates the sparsity of effective alphas.

The infinite search space and sparse effective solutions lead to the following specific issues:

\begin{itemize} 
\item \textbf{Inefficient Search Process:} 
The infinite search space, sparse effective alphas, and financial data complexity make traditional GP inefficient in finding stock selection factors. 
The "code bloat" problem also further hampers search efficiency \cite{Soule2002}. 

\item \textbf{Premature Convergence:} 
Due to the scarcity of effective solutions, once a local optimum is found, a dominant gene quickly takes over the population, causing stagnation and high correlation in the results \cite{Gupta2012}. 

\item \textbf{ Poor interpretability:} 
Random search in large spaces can find alphas that perform well numerically but are too complex or difficult to interpret, making them hard to apply in quantitative trading.
\end{itemize}

Some related works have attempted to address these issues. 
For example, \citet{Gupta2012} proposed several methods to mitigate premature convergence, such as \textit{Warm Start} and \textit{Replacement Method}. 
\citet{Zhang2020} introduced a hierarchical structure to improve the traditional GP framework.
\citet{Zhaofan2022} suggested penalizing the correlation between newly generated alphas and existing ones in the pool to reduce the risk of high correlation. 
Additionally, \citet{Zhang2020} also proposed a \textit{PCA-QD} method, which used principal component correlation as a substitute for traditional correlation measures to reduce computational overhead.
 
In this paper, we propose a new GP framework with carefully chosen initialization and structural constraints, which helps mitigate the aforementioned challenges and achieves promising results in empirical analysis. The main contributions of this paper are as follows:

\begin{itemize}
\item \textbf{Providing a Strong Starting Point:} 
We improve GP's initialization by offering a well-chosen starting point, reducing inefficiencies and enhancing search performance. 
We also propose a simple principle for selecting the initial point.

\item \textbf{Constraining the Tree Structure:} 
We restrict GP's search to a predefined tree structure, supported by a new designed crossover operator that swaps subtrees at equivalent positions within the same structure. 
This approach also ensures the interpretability of the alphas.

\item \textbf{Addressing Premature Convergence and High Correlation:} 
We mitigate premature convergence by avoiding duplicate individuals and reduce result correlation through parallel alpha mining across multiple structures.

\item \textbf{Empirical Validation in the Chinese Stock Market:} Using Chinese stock market data, we demonstrate the proposed framework's superiority over traditional GP and benchmarks in factor mining (IC analysis) and factor returns (Portfolio).

\end{itemize}

As mentioned above, our framework provides a strong starting point and restricts the alpha search to a given structure. This means initializing the genetic search from a well-designed starting point, while limiting the search to a promising space around the given starting point. 
Therefore, we believe that, in a sense, our framework represents another novel approach to providing a warm start for GP. 
Therefore, in the following sections, we will sometimes refer to our framework as the \textbf{Warm Start GP} framework.

\section{Framework Details}
\subsection{Motivation}
After a period of exploring traditional GP, we quickly shift our research focus to identifying effective search spaces. 
With a set of alpha factors already proven to be relatively effective, we aim to construct a potential space centered around these alphas and analyze the distribution of effective alphas within this smaller search space. 
If the distribution of effective alphas in this space is proven denser than in the full search space, we would have identified the desired \textit{effective search space}. 
Building on this, we could establish not only a framework for mining effective alphas but also a \textit{alpha enhancer}, effectively strengthening the given initial alphas. 
However, a crucial question remains: how should we define this potential space centered around a specific alpha?

To address this question and make our results more relevant to the Chinese stock market, we review nearly all recent factor construction reports from leading Chinese securities firms.  
Our approach is inspired by the practices these firms follow when manually constructing alphas. 
Observe the following four alphas:

\begin{itemize}
\item $Alpha_1$: Rank the \textbf{\textit{daily intraday returns}} of the past 20 days by \textbf{\textit{the same day's turnover rate}}, and average the top 5.
\item $Alpha_2$: Rank the \textbf{\textit{daily intraday turnover rates}} of the past 20 days by \textbf{\textit{the next day's overnight turnover rate}}, and average the top 5.
\item $Alpha_3$: Rank the \textbf{\textit{overnight returns}} of the past 20 days by \textbf{\textit{the previous day's turnover rate}}, and average the top 5.
\item $Alpha_4$: Rank the \textbf{\textit{daily intraday turnover rates}} of the past 20 days by \textbf{\textit{the same day's intraday returns}}, and average the top 5."
\end{itemize}

The four alphas mentioned above are all manually constructed by securities firms and have proven effective for actual trading. 
From these alphas, a clear construction pattern can be extracted: 
\begin{itemize}
\item Rank the \textbf{\textit{Data}}$_1$ of the past 20 days by \textbf{\textit{Data$_2$}}, and average the top 5. 
\end{itemize}
This pattern can be further generalized as: 
\begin{itemize}
\item Rank the \textit{\textbf{Data$_1$}} of the past \textbf{\textit{D}} days by \textbf{\textit{Data$_2$}}, and select a \textbf{\textit{subset}} of values to compute a \textbf{\textit{statistical measure}}. 
\end{itemize}

\begin{figure}[ht]
\vskip 0.2in
\begin{center}
\centerline{\includegraphics[width=0.6\columnwidth]{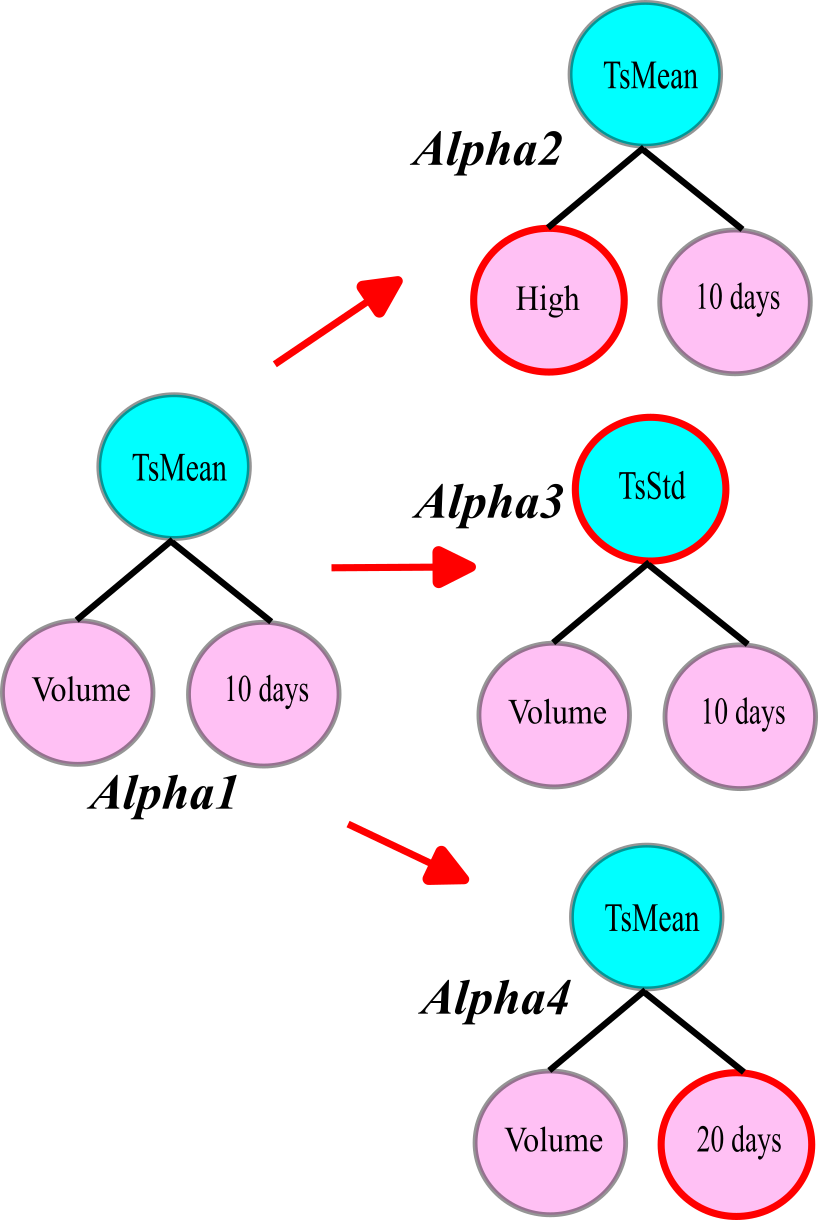}}
\caption{Alphas2, 3, and 4 are alphas that share the same structure as the effective Alpha1.}
\label{fig:Factors in same structure}
\end{center}
\vskip -0.2in
\end{figure}

The \textit{“subset”} can be the top 5, top 10, or bottom 10, and the \textit{“statistical measure”} can be the mean, standard deviation, median, or even mode. 
Mapping this fixed pattern to formula-based or tree-structured alphas corresponds to a predefined alpha structure or a fixed tree structure.
Examples like these are abundant. 
Inspired by this observation, we propose the following Hypothesis \ref{hyp1:structure-effectiveness}: 

\begin{hypothesis}
An alpha's effectiveness comes not only from its variables and functions but also from its underlying structure. 
\label{hyp1:structure-effectiveness}
\end{hypothesis}
Put more intuitively, an effective alpha structure increases the likelihood of identifying impactful alphas, which means maintaining the structure of an effective alpha while modifying its data or functions increases the chances of discovering a new effective alpha compared to random construction. 
The Figure \ref{fig:Factors in same structure} illustrates the concept of identical structures in tree-based formulas. 
All four alphas share a uniform structure consisting of one root node and two leaf nodes with a depth of 1. Under the Hypothesis \ref{hyp1:structure-effectiveness}, if the structure of Alpha1 is effective, Alphas2, 3, and 4 are more likely to be effective as well, as they share the same structure.

\subsection{Identify an Effective Alpha Structure}
Hypothesis\ref{hyp1:structure-effectiveness} suggests that if we have an effective alpha structure, we can more efficiently identify effective stock selection factors within this structure. 
This is an encouraging discovery, but a new question arises: how do we define and identify an effective alpha structure? 
Undoubtedly, we could rely on economists or financial engineers to design such a structure for us, as some securities firms are doing. 
However, not every investor has such extensive investment experience or economic knowledge. 
So it is essential to develop a method for efficiently and intuitively identifying effective alpha structures, ensuring the framework's accessibility and usability for a broader range of investors.

To address this challenge, we propose Hypothesis \ref{hyp2:factor-effectiveness}: 
\begin{hypothesis}
An effective alpha is often characterized by an effective structure.
\label{hyp2:factor-effectiveness}
\end{hypothesis}

This hypothesis offers investors a simple and practical principle for selecting effective structures: 

\textbf{If an alpha is validated as effective, its underlying structure is also effective and can be used to construct an effective search space.}

In other words, investors can simply identify an effective alpha and adopt its structure. 
Moreover, alphas that were highly effective in the past but are temporarily ineffective can also be considered. 
While their theoretical basis may still be under investigation, our experiments show that these structures are often valid, leading us to speculate that short-term inefficacy may be due to certain data or functions not accurately reflecting the current market, rather than a failure of the factor structure itself.
Finally, while any eligible alpha can be selected, it's preferable to choose those with simple structures, clear functions, and strong interpretability, given computational constraints.

\begin{figure}[ht]
\vskip 0.2in
\begin{center}
\centerline{\includegraphics[width=\columnwidth]{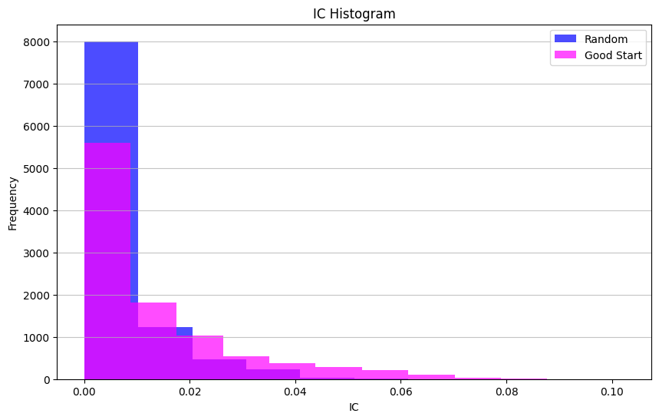}}
\caption{The blue segment represents fully random results, previously shown in Figure\ref{fig:Random Factor1}, while the purple segment reflects the results obtained under the given structural constraints.}
\label{fig:Random Factor2}
\end{center}
\vskip -0.2in
\end{figure}

To validate Hypothesis \ref{hyp1:structure-effectiveness} and Hypothesis \ref{hyp2:factor-effectiveness}, we conduct an additional experiment.
Specifically, we select an alpha from the Alpha101
\footnote{The majority of the alphas in Alpha101 have been validated as effective by both the U.S. and Chinese markets. The alpha selected for our experiment is Alpha33.} 
set and randomly generate another 10,000 alphas on dataset consistent with the experiment in Figure \ref{fig:Random Factor1}, within the structural constraints of this alpha. 
The results are then compared with those in Figure \ref{fig:Random Factor1}, as illustrated in Figure\ref{fig:Random Factor2}.  
The findings reveal a substantial increase in the density of effective alphas within the constrained search space. 
The proportion of alphas with $IC > 0.03$ exceeded $13\%$, more than tripling the density observed in the unconstrained space. 
This outcome confirms Hypothesis \ref{hyp1:structure-effectiveness}. 
Furthermore, since the structure used in this experiment originated from a known effective alpha, Hypothesis \ref{hyp2:factor-effectiveness} is also validated.

\subsection{Warm Start GP Framework}
Based on the two hypotheses outlined above, this paper proposes a framework for mining stock selection factors using genetic programming with given initialization and structural constraints. 
The proposed framework provides GP with a strong starting point, which is a single individual rather than the traditional initial population in traditional GP. 
This initial individual not only provides a good starting point for the search, accelerating convergence, but also offers an effective alpha structure. 
The selection principle for this initial individual is guided by Hypothesis \ref{hyp2:factor-effectiveness}. 
Starting from this initial individual, our framework restricts the search to the structure possessed by this individual, continuing until the termination condition is met or the maximum number of iterations is reached, after which the best individual found within the structure is output.

Moreover, our framework is not only a robust alpha mining framework but also an excellent alpha enhancing framework. 
In fact, within our framework, these two concepts are equivalent. 
We provide GP with a good starting point as a warm start, which not only locates the initial position but also locks in a small search space. 
From one perspective, this approach allows for quick identification of an effective search space, thus speeding up the alpha search process. 
From another perspective, if we focus on the given starting point - the chosen effective alpha - itself, the work we are actually doing is enhancing the given alpha. 
Returning to Figure \ref{fig:Factors in same structure}, you will notice that Alphas2, 3, and 4 are all attempts to enhance Alpha1. 
In reality, our framework operates in precisely this manner: it aims to improve the performance of a given alpha by making modifications that do not disrupt the structure of the alpha.

\begin{figure}[ht]
\vskip 0.2in
\begin{center}
\centerline{\includegraphics[width=0.75\columnwidth]{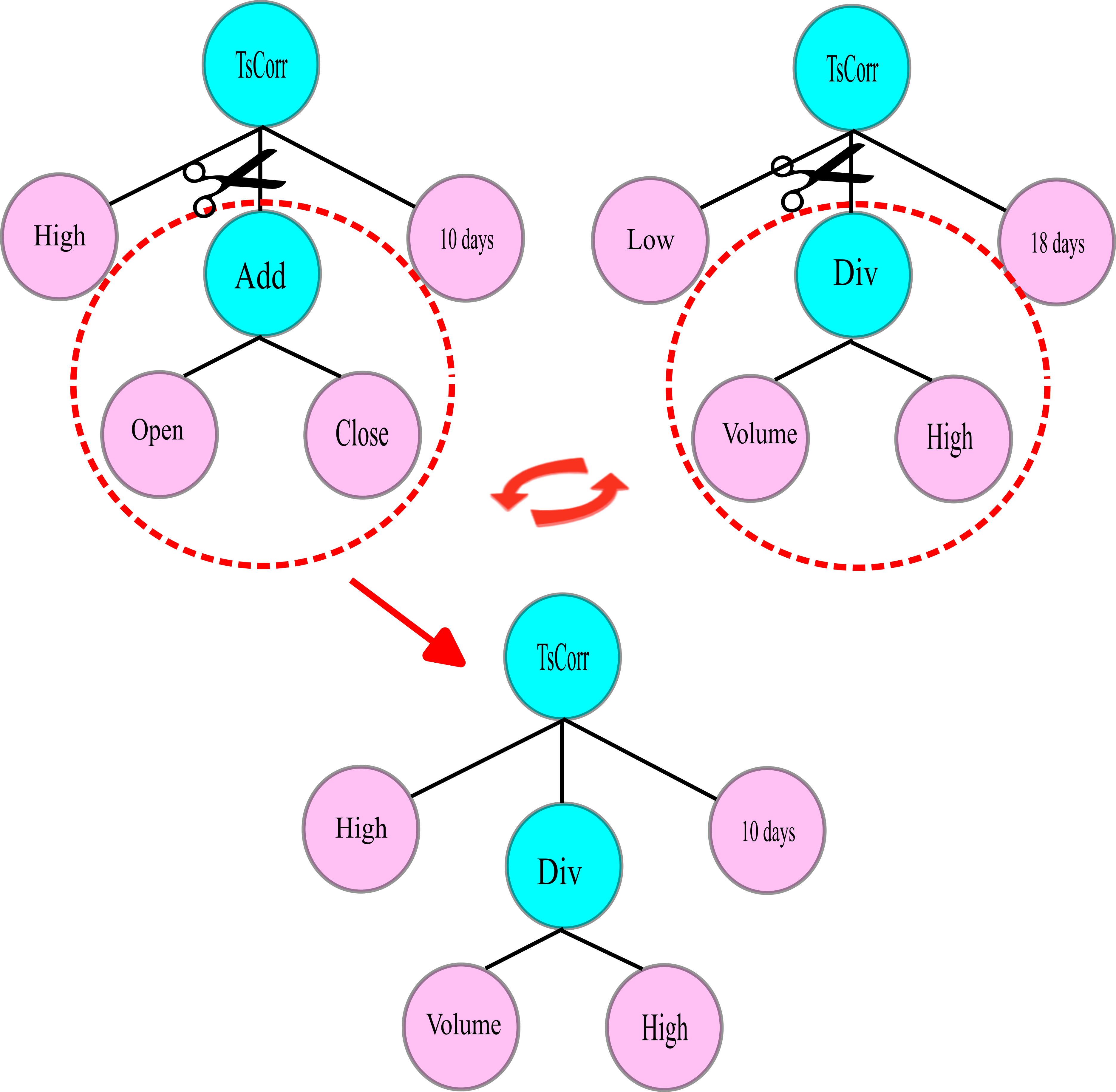}}
\caption{The restricted crossover: only allows alphas with the same structure to exchange subtrees at the same positions, ensuring that the alpha structure remains unchanged.}
\label{fig:Constrained Crossover}
\end{center}
\vskip -0.2in
\end{figure}

\begin{algorithm}[ht]  
\caption{Warm Start GP Framework}
\label{Warm Start GP Framework}
\begin{algorithmic}[1]
\INPUT Fitness function $f(\cdot)$, Population size $n_{\text{pop}}$, Mutation rates $P=(p_{\text{crossover}}, p_{\text{point}})$, other initial params $Params_{init}$, Stock data $\textbf{X}$, Forward return $\textbf{Y}$, initial alpha $alpha_{init}$
\OUTPUT An enhanced alpha
\STATE $t \gets 0$
\STATE $Pop(t) \gets alpha_{init}$ 
\STATE EvaluatePopulation($Pop(t)$)
\WHILE{not terminate}
    \STATE $Pop(t + 1) \text{~~~Insert~~~} \text{argmax } f(Pop(t))$
    \WHILE{$\text{len}(Pop(t+1)) < n_{\text{pop}}$}
        \IF{$t + 1 == 1$}
            \STATE $Mutation$ $\gets$ PointMutation 
        \ELSE
            \STATE $Mutation$ $\gets$ Choose method randomly($P$)
        \ENDIF
        \IF{$Mutation$ == Crossover}
            \STATE $P_1 \gets \text{Tournament}(Pop(t))$
            \STATE $P_2 \gets \text{Tournament}(Pop(t))$
            \STATE $Offspring \gets \text{Crossover}(P_1, P_2)$
        \ELSE
            \IF{$Mutation$ == PointMutation}
                \STATE $P_1 \gets \text{Tournament}(Pop(t))$
                \STATE $Offspring \gets \text{PointMutation}(P_1)$
            \ELSE
                \STATE $P_1 \gets \text{Tournament}(Pop(t))$
                \STATE $Offspring \gets P_1$
            \ENDIF     
        \ENDIF
        \IF{$Offspring \text{~~~} \text{ NOT IN } \text{~~~} Pop(t+1)$}
            \STATE $Pop(t+1) {~~~} \text{Insert} {~~~} Offspring$
        \ENDIF
    \ENDWHILE
    \STATE EvaluatePopulation($Pop(t + 1)$)
    \STATE $t \gets t + 1$
\ENDWHILE
\end{algorithmic}
\end{algorithm}

The Algorithm \ref{Warm Start GP Framework} provides a detailed explanation of how to search for superior alphas within the structure of a given alpha factor. 
The key distinction of the proposed framework compared to traditional GP lies in starting from a known effective alpha and restricting the search to its predefined structure.
To implement this critical distinction, we design a \textbf{Restricted Crossover Operator}, allowing only restricted crossover and point mutation during the crossover and mutation phases, ensuring that the alpha structure remains unchanged throughout the search process. 
The restricted crossover operator supports the exchange of identical subtrees at the same positions between alphas of the same structure, thereby preserving the alpha structure. 
This step is detailed in line 15 of the Algorithm \ref{Warm Start GP Framework} and is illustrated with a simple example of restricted crossover in Figure \ref{fig:Constrained Crossover}. 
In this paper, the effective alpha structures we selected all come from Alpha101.

Since the first generation at $t=0$ contains only a single given individual, crossover and mutation cannot be applied to a single individual. 
Thus, during the construction of the second generation ($t+1 = 1$ in the algorithm), only point mutation is allowed to ensure the generation of a sufficient number of alphas with the same structure but different details. 
For generations with $t>1$, selected parents undergo restricted crossover or point mutation with a given probability, and through the process of selection and elimination, the final population is obtained.

A few points require clarification: We use the tournament selection method to choose parents. In line 5, we ensure the best individual from the previous generation is carried over to prevent performance decline. In line 25, we avoid duplicate individuals in the next generation to prevent rapid gene domination. Finally, the algorithm starts from a single point, but practical applications often use parallel searches from multiple starting points.

\subsection{Why Warm Start GP framework}

In the former Sections, we briefly introduced the benefits of using valid initial alphas and structures. Here, we elaborate on the advantages of the proposed framework:
\begin{itemize}
\item \textbf{Warm Start with Efficient, Finite Search Space: } 
As shown in Figure \ref{fig:Random Factor2}, selecting an effective alpha structure focuses the search on a sub-space that is not only more efficient but also finite. 
An effective and finite search space, combined with a well-defined starting point, enables our framework to demonstrate strong performance in alpha mining tasks.
Additionally, restricting the tree structure completely avoids the "code bloat" problem, improving computational efficiency.

\item \textbf{A Strong Alpha Enhancer:} 
A finite search space ensures that, in theory, the optimal solution under the sapce can be found. 
This means another strength of our framework, as even with other more efficient alpha-mining methods, the proposed framework can serve as an \textbf{"Alpha Enhancer"} to improve the alphas obtained. 

\item \textbf{Reducing Overfitting Risk and improving interpretability:} Traditional GP incline to increase solution depth and length to improve performance, but this can lead to poor out-of-sample results. 
Evidence from \citet{Gupta2012} suggests that restricting the search to a given structure can mitigate this risk.
In addition, constraining the structure of the alphas means we can provide well-defined, interpretable structures and prevent the alphas from becoming unrecognizable. This significantly enhances the interpretability of the resulting alphas.

\item \textbf{Reducing Premature Convergence and High Correlation Issues:} 
The proposed framework prevents the entry of duplicate individuals into the population, reducing the likelihood of domination by a small set of genes. 
Furthermore, since our framework works within a single structure and we expect only one optimal individual, the likelihood of high correlation among individuals across different structures becomes lower. 

\end{itemize}

The advantages of the proposed framework outlined above will be further validated and demonstrated in detail in the next section through empirical analysis.

\section{Numerical Results}
\subsection{Experiment Settings}
To validate the several mentioned advantages of the proposed framework in alpha mining, we conduct experiments on \textbf{Correlation Analysis}, \textbf{IC Analysis}, and \textbf{Trading Performance}.  
We set a 5-day holding period to evaluate the alpha's ability to predict stock returns. 
Starting with 10 effective alphas from Alpha101, we use our framework to generate 10 high-quality alphas and compare the results with both the chosen Alpha101 alphas and those generated by traditional GP.

\textbf{Datasets:} We use the full A-shares data of the Chinese stock market for alpha discovery, with the alpha mining period spanning from 2020.01 to 2021.12. 
The testing period is set from 2022.01 to 2024.10.

\begin{figure}[ht]
\vskip 0.2in
\begin{center}
\begin{minipage}{0.45\textwidth}
  \centering
  \includegraphics[width=\linewidth]{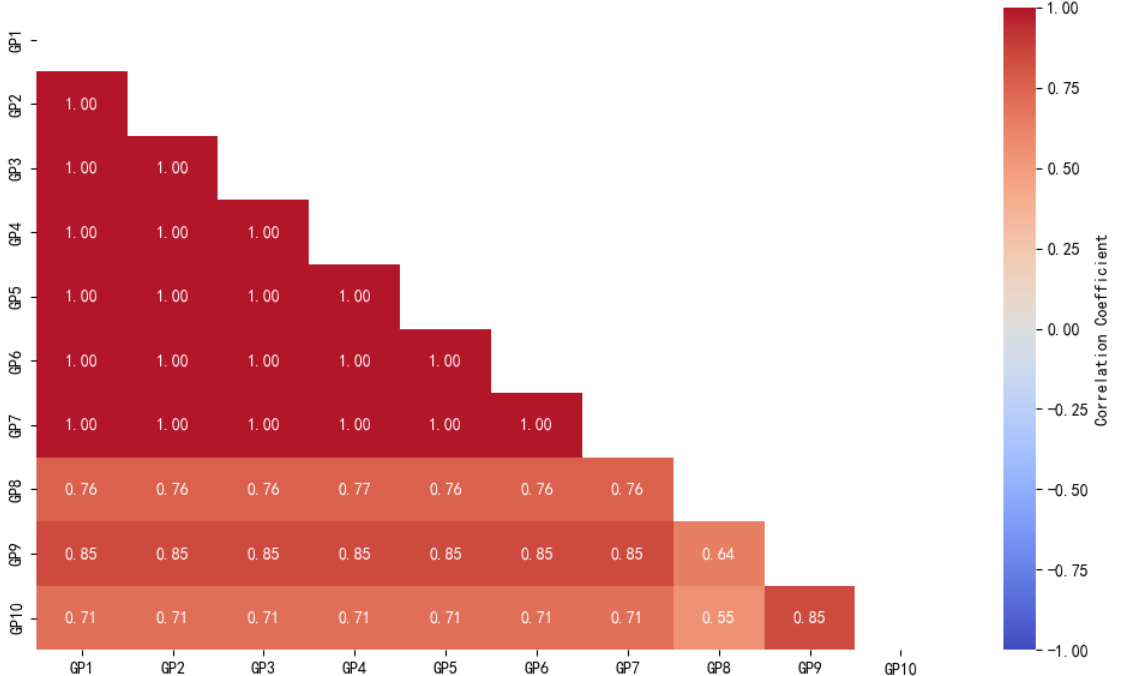}
  \subcaption*{a. GP: average absolute value 0.87} \label{fig:GPcorr}
\end{minipage}\hfill
\begin{minipage}{0.45\textwidth}
  \centering
  \includegraphics[width=\linewidth]{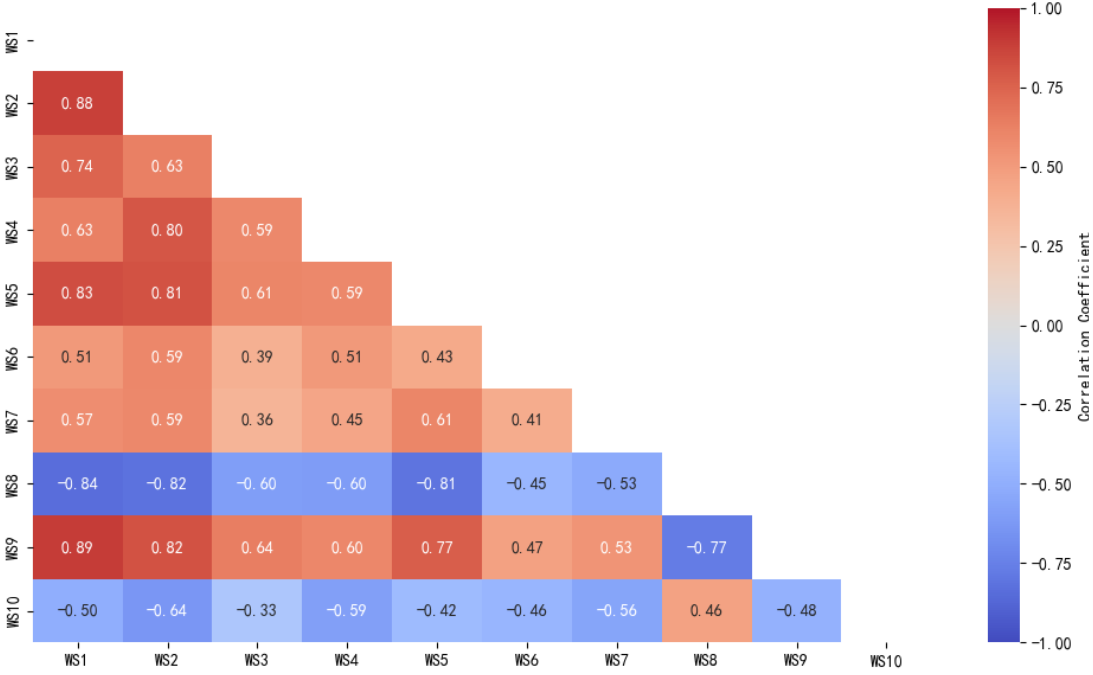}
  \subcaption*{b. Warm Start GP: average absolute value 0.60} \label{fig:WScorr}
\end{minipage}
\caption{Correlation coefficients of GP alphas (a.) and WS alphas (b.).}
\label{fig:corr}
\end{center}
\vskip -0.2in
\end{figure}

\subsection{Correlation Analysis}
We aim for each GP program to generate distinct, high-quality alphas, as highly correlated or identical factors can harm downstream tasks (e.g., multicollinearity). 
Traditional GP often gets stuck in local optima, leading to highly correlated results both within a single run and across multiple runs.
The Figure \ref{fig:corr} below shows Spearman correlations among the 10 best alphas from 10 traditional GP runs, where seven runs even produce identical factors—an outcome we seek to avoid.

The alphas derived from different starting points by the Warm Start GP framework
\footnote{These are referred to as \textbf{WS} alphas in the Figure \ref{fig:corr}}
reduce the average correlation to around 0.6, without producing identical factors, demonstrating its advantage over traditional GP in mitigating high correlation.

\subsection{IC Analysis}

The IC analysis method is commonly used to evaluate alpha effectiveness. 
The evaluation metrics we use are IC, ICIR, RankIC, and RankICIR, where RankIC replaces the Pearson correlation with the Spearman correlation in calculating IC, and ICIR is the ratio of the mean IC to its standard deviation, reflecting alpha significance and stability:

\[
ICIR = \frac{IC}{\text{std}(\text{PearsonCorr}(\textbf{a}_t, \textbf{r}_{t})}  \]
\[
RankIC = \frac{1}{T} \sum_{t=1}^{T} \text{SpearmanCorr}(\textbf{a}_t, \textbf{r}_{t})
\]
\[
RankICIR = \frac{RankIC}{\text{std}(\text{SpearmanCorr}(\textbf{a}_t,\textbf{r}_{t})} 
\]
Larger values of these metrics are better. 

In the IC analysis, we first demonstrate how the proposed framework enhances the initial alphas from Alpha101 set.  
We compare the IC metrics before and after enhancement and the results are presented in Table \ref{table: enhanceIC}.
We also perform an IC analysis on the top 10 alphas generated by traditional GP in Table \ref{table: GPIC} to assess whether our framework provides an improvement over traditional GP in the alpha mining task for quantitative investment. 
To avoid duplication, we select the second-best alpha when there are repeats among the top 10 GP alphas, ensuring each GP program contributed a unique alpha. 

\begin{table*}[t]
\caption{IC analysis results of the selected starting points from Alpha101 (A101), with "$\#$" indicating its code in the Alpha101 set, and the corresponding alphas enhanced by our proposed framework (WS), sorted by the out-of-sample IC of the WS alphas.
The results outside (inside) the brackets indicate out-of-sample (in-sample) results, with the sample division being consistent with the mining and testing period split in Experiment Settings.}
\label{table: enhanceIC}
\vskip 0.1in
\begin{center}
\setlength{\tabcolsep}{5.5pt}
\begin{small}
\begin{sc}
\begin{tabular}{lcccccccc}
\toprule
$\#$    
& \multicolumn{2}{c}{$IC$} 
& \multicolumn{2}{c}{$ICIR$} 
& \multicolumn{2}{c}{$RankIC$} 
& \multicolumn{2}{c}{$RankICIR$} \\
\midrule
$ $ & A101 & WS & A101 & WS & A101 & WS & A101 & WS\\
\cmidrule(lr){2-3} \cmidrule(lr){4-5} \cmidrule(lr){6-7} \cmidrule(lr){8-9}
$a_{025}$ & 0.011(0.011) & \textbf{0.060}(0.035) & 0.11(0.14) & \textbf{0.46}(0.24) & 0.011(0.013) & \textbf{0.091}(0.067) & 0.09(0.13) & \textbf{0.59}(0.38) \\
$a_{067}$ & 0.000(0.000) & \textbf{0.056}(0.033) & 0.00(0.00) & \textbf{0.47}(0.24) & 0.000(0.000) & \textbf{0.081}(0.066) & 0.00(0.00) & \textbf{0.57}(0.42) \\
$a_{047}$ & 0.015(0.013) & \textbf{0.051}(0.035) & 0.14(0.16) & \textbf{0.54}(0.33) & 0.005(0.011) & \textbf{0.085}(0.073) & 0.03(0.11) & \textbf{0.76}(0.61) \\
$a_{005}$ & 0.013(0.003) & \textbf{0.051}(0.034) & 0.17(0.04) & \textbf{0.43}(0.26) & 0.006(0.002) & \textbf{0.082}(0.069) & 0.06(0.02) & \textbf{0.56}(0.45) \\
$a_{090}$ & 0.004(0.013) & \textbf{0.050}(0.032) & 0.04(0.16) & \textbf{0.45}(0.26) & 0.007(0.011) & \textbf{0.076}(0.064) & 0.06(0.11) & \textbf{0.57}(0.47) \\
$a_{008}$ & 0.018(0.013) & \textbf{0.048}(0.033) & 0.18(0.18) & \textbf{0.42}(0.25) & 0.022(0.019) & \textbf{0.089}(0.077) & 0.18(0.23) & \textbf{0.67}(0.49) \\
$a_{040}$ & 0.036(0.028) & \textbf{0.042}(0.034) & 0.41(0.37) & \textbf{0.48}(0.43) & 0.069(0.057) & \textbf{0.076}(0.068) & 0.67(0.63) & \textbf{0.78}(0.73) \\
$a_{011}$ & 0.001(0.004) & \textbf{0.042}(0.035) & 0.02(0.06) & \textbf{0.42}(0.33) & 0.006(0.000) & \textbf{0.058}(0.056) & 0.06(0.00) & \textbf{0.46}(0.45) \\
$a_{018}$ & 0.029(0.022) & \textbf{0.037}(0.041) & 0.30(0.25) & \textbf{0.36}(0.36) & 0.041(0.037) & \textbf{0.075}(0.080) & 0.36(0.36) & \textbf{0.62}(0.57) \\
$a_{101}$ & 0.018(0.011) & \textbf{0.033}(0.033) & 0.16(0.14) & \textbf{0.24}(0.27) & 0.019(0.016) & \textbf{0.066}(0.066) & 0.15(0.18) & \textbf{0.42}(0.49) \\
\midrule
$avg$ & 0.015(0.012) & \textbf{0.047}(0.034) & 0.15(0.15) & \textbf{0.43}(0.30) & 0.019(0.017) & \textbf{0.078}(0.068) & 0.17(0.18) & \textbf{0.60}(0.51) \\
\bottomrule
\end{tabular}
\end{sc}
\end{small}
\end{center}
\vskip -0.1in
\end{table*}

According to the results in Table \ref{table: enhanceIC}, the proposed framework significantly improve the performance of the selected Alpha101 alphas, with the average in-sample IC increasing from $1\%$ to over $3\%$, and the average RankIC rising to $\textbf{6.8}\%$. 
The result also shows that this improvement is maintained out-of-sample. 
While the original alphas show little difference in performance between in- and out-of-sample, the enhanced alphas perform even better out-of-sample, with an average IC of $\textbf{4.7}\%$ and RankIC of $\textbf{7.8}\%$. 
These results confirm that our framework enhances alphas from a solid starting point, with clear improvements both in- and out-of-sample.

\begin{table}[t]
\caption{Out-of-(in-) Sample IC performances of traditional GP alphas, sorted by the out-of-sample IC of the GP alphas}
\label{table: GPIC}
\vskip 0.1in
\begin{center}
\setlength{\tabcolsep}{3pt}
\begin{small}
\begin{sc}
\begin{tabular}{lcccr}
\toprule
$\#$    & $IC$ & $ICIR$ & $RankIC$ & $RankICIR$ \\
\midrule
1  &  0.056 (0.034)  &  0.48 (0.26)  &  0.085 (0.066)  &  0.60 (0.44)  \\
2  &  0.055 (0.032)  &  0.47 (0.25)  &  0.084 (0.067)  &  0.59 (0.44)  \\
3  &  0.039 (0.030)  &  0.33 (0.26)  &  0.070 (0.072)  &  0.53 (0.61)  \\
4  &  0.036 (0.037)  &  0.28 (0.30)  &  0.078 (0.079)  &  0.53 (0.55)  \\
5  &  0.036 (0.040)  &  0.26 (0.31)  &  0.073 (0.077)  &  0.45 (0.52)  \\
6  &  0.034 (0.038)  &  0.30 (0.35)  &  0.072 (0.077)  &  0.53 (0.60)  \\
7  &  0.030 (0.038)  &  0.22 (0.30)  &  0.063 (0.076)  &  0.40 (0.52)  \\
8  &  0.028 (0.037)  &  0.23 (0.33)  &  0.067 (0.077)  &  0.46 (0.58)  \\
9  &  0.028 (0.036)  &  0.21 (0.27)  &  0.064 (0.074)  &  0.39 (0.47)  \\
10 &  0.022 (0.038)  &  0.18 (0.35)  &  0.055 (0.070)  &  0.37 (0.53)  \\
\midrule
$avg$ & 0.036 (0.036)  &  0.30 (0.30)  &  0.071 (0.074)  &  0.48 (0.52)  \\
\bottomrule
\end{tabular}
\end{sc}
\end{small}
\end{center}
\vskip -0.1in
\end{table}

Compared to traditional GP, the proposed framework also demonstrates a significant advantage in out-of-sample alpha performance. 
The alphas identified by our framework have an average out-of-sample IC that is more than $\textbf{1}\%$ higher than those from traditional GP, and the average RankIC has increased $\textbf{1}\%$, let alone a more pronounced advantage in ICIR and RankICIR.
Considering that our framework does not have an advantage over traditional GP in-sample, the comparison results in Tables \ref{table: enhanceIC} and \ref{table: GPIC} demonstrate that the alphas discovered by our framework not only outperform those from traditional methods, but also alleviate the overfitting problem commonly seen in traditional GP.

\subsection{Trading Performance}
IC analysis is one aspect of measuring the effectiveness of an alpha. 
For actual trading, the alpha's ability to generate excess returns is more critical. 
We use the backtesting results of the alphas to assess their ability to capture excess returns. 

\begin{itemize}
\item \textbf{Backtesting Setup:}
To avoid using future information in any trade, we split the test period into two segments: 2022.1 to 2022.12 for training the alpha model, and 2023.1 to 2024.10 for forecasting future returns by the trained alpha model. 
To simulate real trading conditions, we apply the following restrictions: stocks cannot be bought if hitting the daily limit-up, and cannot be sold if hitting the limit-down or being suspended. 
Transaction costs are 0.6\textperthousand.
\item \textbf{Investment Strategy:}
The holding period is set to 5 trading days.
At each rebalancing, we invest in the top-ranked stocks with equally distributed capital. 
If a stock cannot be bought, its allocation is redistributed to the remaining top-ranked stocks. 
We set different holding sizes: 10, 30, and 100 stocks.
The buy and sell prices are set to be the VWAP of the day. 
Furthermore, we set a rule that if a stock is flagged with an ST or *ST warning, we will no longer hold that stock, which is a common risk-aversion measure in real-world trading.
\item \textbf{Alpha Model:}
For simplicity, we opt for a linear regression model to fit the alpha factors.
The predicted value on day \textit{T} is the forecasted cumulative return of the VWAP from day \textit{T+1} to day \textit{T+6}. 
Thus, when executing trades on day \textit{T}, stock rankings are based on the predicted returns from day \textit{T-1}.
\item \textbf{Benchmark:}
 We construct alpha models for the alphas generated by our framework, the selected Alpha101 alphas, and traditional GP alphas using data from 2022. 
 Portfolios are then built based on their respective predictions after 2022. Additionally, we select the CSI 300 (hs300) Index, CSI 500 (zz500) Index, CSI 1000 (zz1000) Index, and the CSI All (zz$\_$all) Index as benchmarks representing the market level.
\end{itemize}

\begin{figure}[ht]
\vskip 0.2in
\begin{center}
\begin{minipage}{0.45\textwidth}
  \centering
  \includegraphics[width=\linewidth]{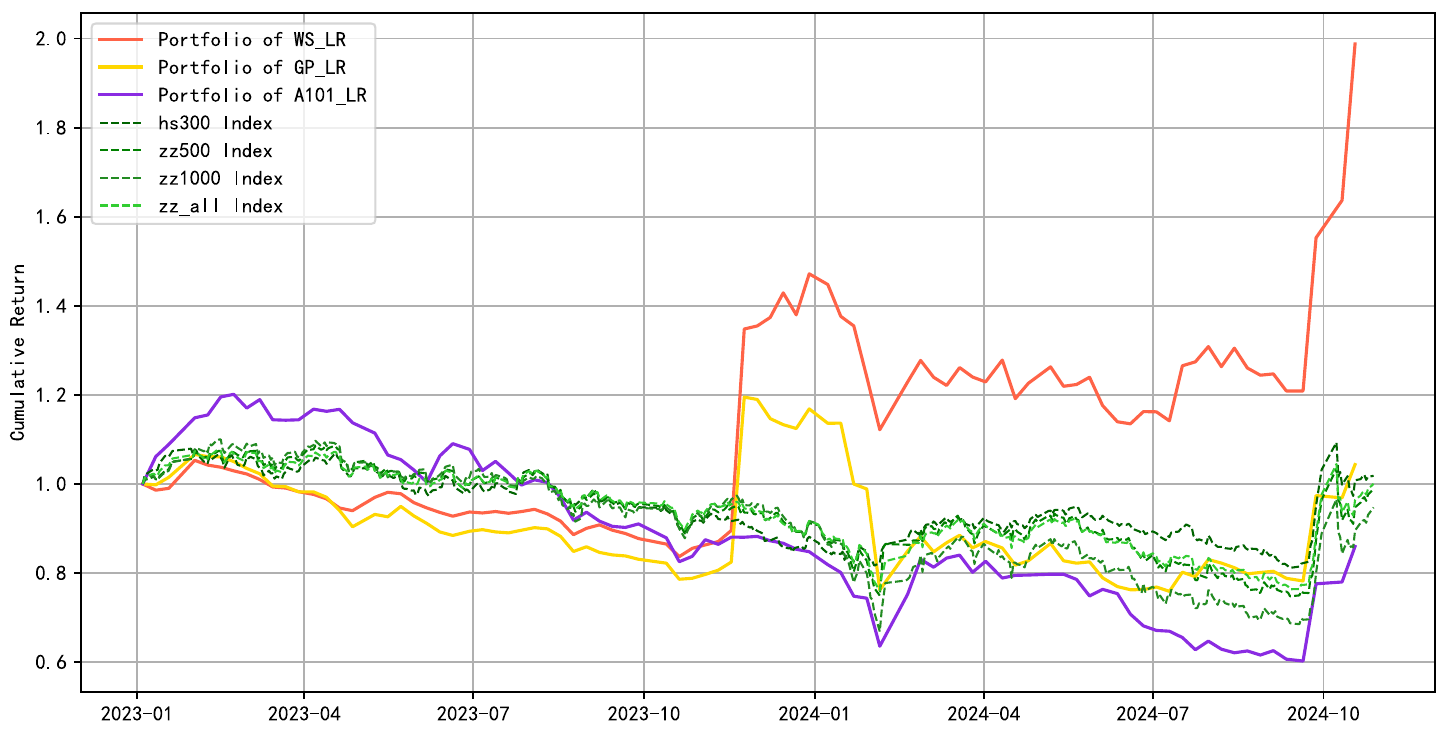}
  \subcaption*{a. Portfolio performance for holding 10 stocks.} \label{fig:portfolio10}
\end{minipage}\hfill
\begin{minipage}{0.45\textwidth}
  \centering
  \includegraphics[width=\linewidth]{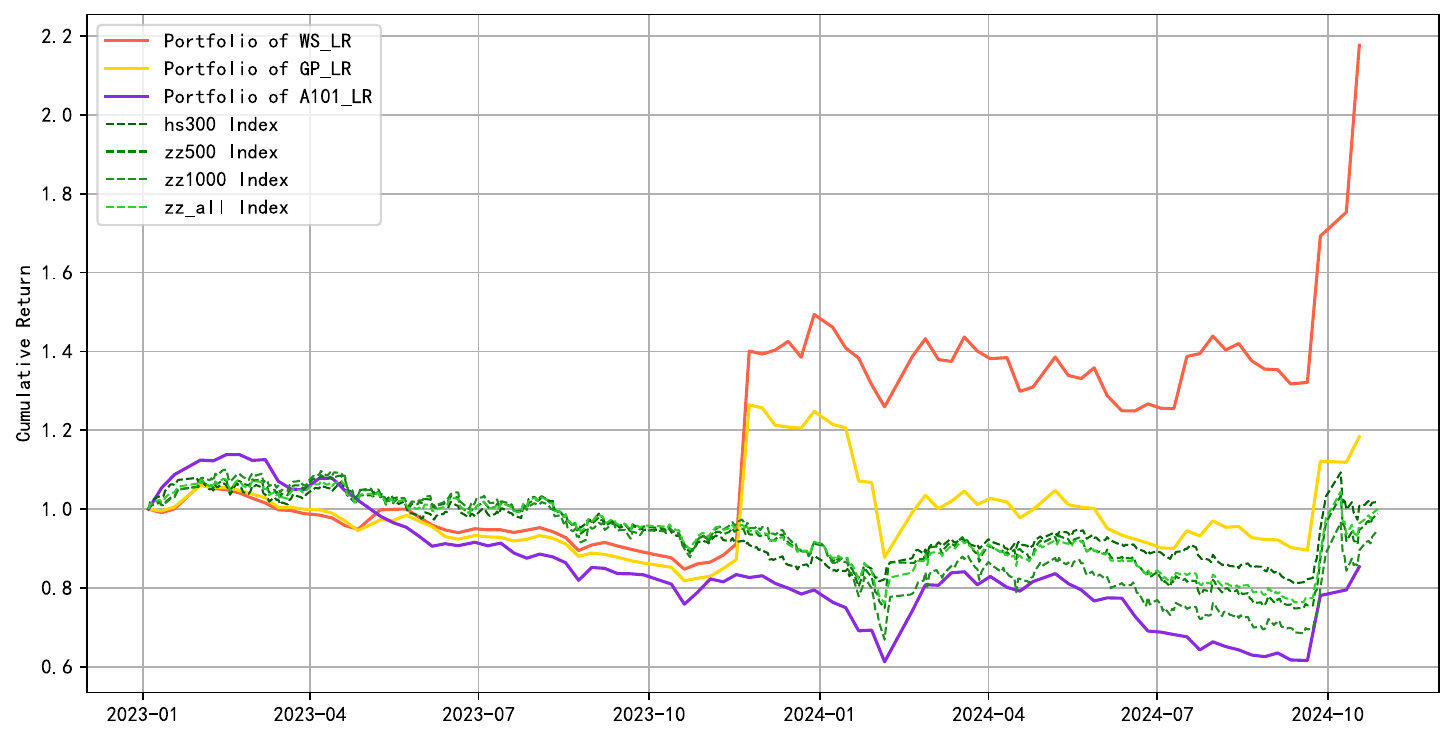}
  \subcaption*{b. Portfolio performance for holding 30 stocks.} \label{fig:portfolio30}
\end{minipage}\hfill
\begin{minipage}{0.45\textwidth}
  \centering
  \includegraphics[width=\linewidth]{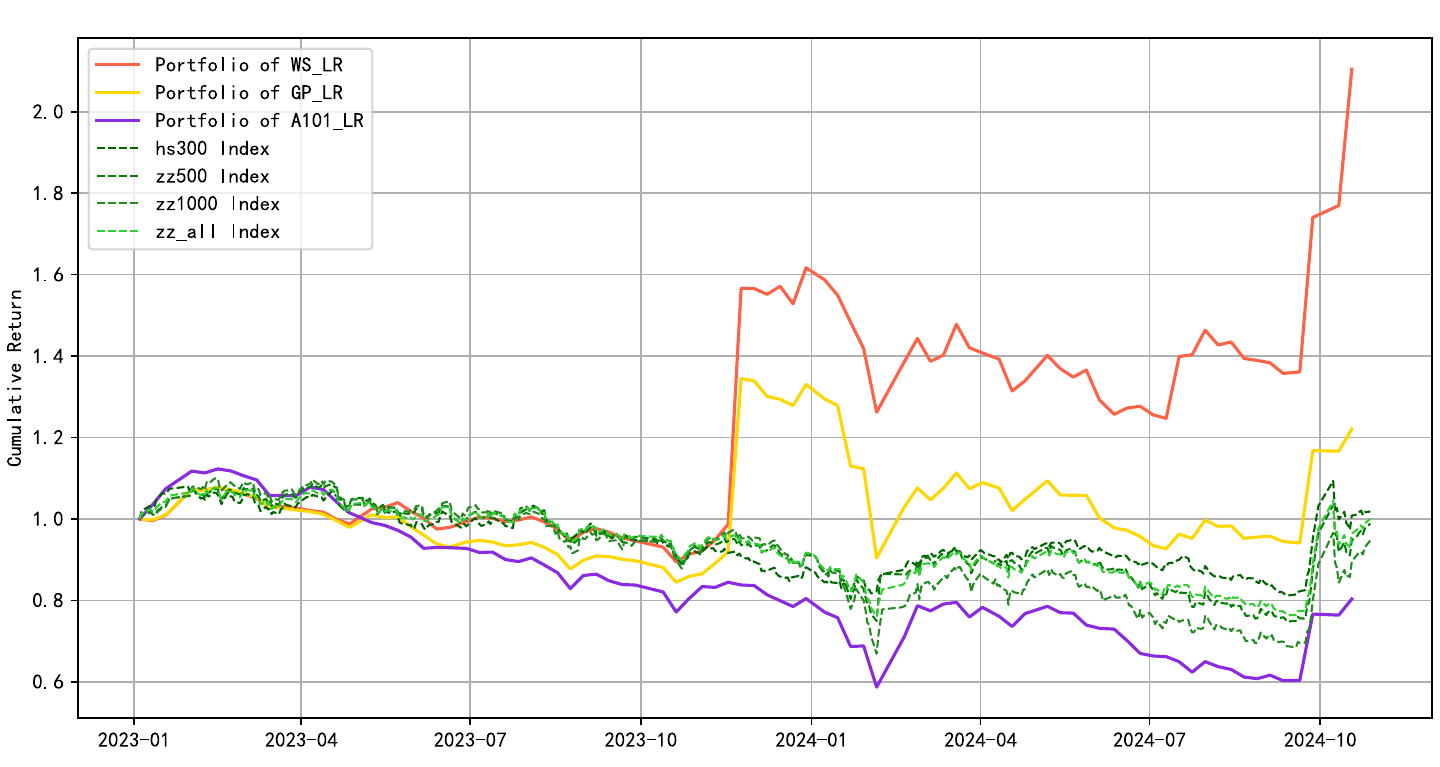}
  \subcaption*{c. Portfolio performance for holding 100 stocks.} \label{fig:portfolio100}
\end{minipage}
\caption{Portfolio performance under different methods and different holding sizes}
\label{fig:portfolio}
\end{center}
\vskip -0.2in
\end{figure}

According to the results in Figure \ref{fig:portfolio}, regardless of the number of stocks held, the model constructed based on the alphas extracted from the proposed framework consistently outperforms the market, the Alpha101 alphas before enhancement, and the traditional GP alphas, achieving the highest portfolio returns.
We also calculated the annualized return (AR) and Sharpe ratio (SR) of the portfolios constructed by the three alpha models. 
The calculation formula of SR is:
\[
\text{SR} = \frac{{Ret}_{P} - {Ret}_f}{\sigma_p}
\]
where ${Ret}_{P}$ is the annualized return of the portfolio, ${Ret}_f$ is the risk-free rate, which is set to 0 in this paper, and $\sigma_p$ is the annualized volatility of the portfolio.

The numerical results of AR and SR in Table \ref{table: ARSR} indicate that the alphas extracted by our framework achieve substantial excess returns far surpassing other benchmarks. 
The annualized return exceeds \textbf{50}$\%$ under both 30 and 100 holding sizes, with the Sharpe ratio exceeding \textbf{1.0} under the 30 holding size. 
This suggests that the alphas identified by the proposed framework are not only superficially strong in terms of IC performance, but they also possess significant investment potential, capable of guiding investors in making profitable investment decisions.

\begin{table}[t]
\caption{AR and SR results of backtests and comparisons with baselines}
\label{table: ARSR}
\vskip 0.1in
\begin{center}
\setlength{\tabcolsep}{3pt}
\begin{small}
\begin{sc}
\begin{tabular}{lrrrrrr}
\toprule
$\#$    
& \multicolumn{2}{c}{$Size = 10$} 
& \multicolumn{2}{c}{$Size = 30$} 
& \multicolumn{2}{c}{$Size = 100$} \\
\midrule
$ $        &AR & SR & AR & SR & AR & SR\\
\cmidrule(lr){2-3} \cmidrule(lr){4-5} \cmidrule(lr){6-7} 
$A101\_LR$ &-0.083 &-0.231 &-0.087 &-0.253 &-0.118 &-0.340 \\
$GP\_LR$   &0.024  &0.052  &0.102  &0.221  &0.122  &0.257 \\
$WS\_LR$   &\textbf{0.484}  &\textbf{0.937}  &\textbf{0.564}  &\textbf{1.059}  &\textbf{0.534}  &\textbf{0.959} \\
\bottomrule
\end{tabular}
\end{sc}
\end{small}
\end{center}
\vskip -0.1in
\end{table}

\section{Conclusion}
We identify several key challenges in traditional GP for alpha factor discovery: the vast search space and the sparsity of effective solutions.
We find that GP performs better when focusing on promising regions rather than random searching, so we propose a new GP framework with carefully chosen initialization and structural constraints, thereby forcing GP to focus the search on more promising areas, which is motivated by the alpha searching practice and aims to boost the efficiency of such a process.
Our analysis of 2020-2024 Chinese stock market data shows that the proposed framework yields superior out-of-sample prediction results and higher portfolio returns than the benchmark.

There are several directions for further research. 
For example,  we employ a simple linear regression model to aggregate the alphas in our work. 
However, more complex machine learning or deep learning models could be used to enhance the factor aggregation process. 
Designing specialized models for alpha synthesis is a significant area of future work. 
Additionally, the computational cost of GP presents a serious challenge during our experiment. 
To date, there has been limited rigorous research on the time consumption of GP in the alpha selection process. 
We believe this is a promising direction for further investigation, such as analyzing which steps consume the most and identifying areas where efficiency can be improved. 
Addressing these challenges would be instrumental in promoting the broader application of GP-based methods in alpha selection.

\bibliographystyle{icml2021}
\bibliography{main}


\end{document}